\documentstyle[emulateapj]{article}

\lefthead{Alonso-Herrero, Rieke \& Rieke}
\righthead{Massive star formation in LIRGs}

\begin{document}

\title{Massive  star formation in Luminous Infrared Galaxies: 
Giant H\,{\sc ii} regions 
and their relation to super star clusters\footnote{Based on observations with the NASA/ESA Hubble Space
Telescope, obtained from the data Archive at the Space Telescope Science
Institute, which is operated by the Association of Universities for
Research in Astronomy, Inc., under NASA contract NAS 5-26555. }}

\author{Almudena Alonso-Herrero, George H. Rieke \& Marcia J. Rieke}
\affil{Steward Observatory, The University of Arizona, Tucson, AZ 85721, USA}

\begin{abstract}
We have used 
{\it HST}/NICMOS broad-band (at $1.6\,\mu$m) and narrow-band 
Pa$\alpha$ ($\lambda_{\rm rest}= 1.87\,\mu$m) images to identify 
star clusters  and H\,{\sc ii} regions  
respectively in a sample of 8 luminous infrared galaxies (LIRGs). 
These observations have 
revealed the presence of a large population of super star clusters and 
bright H\,{\sc ii} regions. A significant fraction of the H\,{\sc ii}
regions shows H$\alpha$ luminosities above 
that  of 30 Doradus, the prototypical giant H\,{\sc ii} region. 
The  excess of extremely luminous H\,{\sc ii} regions in LIRGs 
has been confirmed by comparison with normal 
galaxies observed at similar spatial resolutions.   
Despite the large numbers of identified star clusters and 
H\,{\sc ii} regions in LIRGs, we only find a small fraction of coincidences, 
between 4\% and 30\% of the total number of detected 
sources. Using evolutionary 
synthesis models we have reproduced the relative fractions of 
young H\,{\sc ii} regions, intermediate and old star clusters
observed in Arp~299 and the central region 
NGC~3256 using a Salpeter IMF and instantaneous star 
formation. 
H\,{\sc ii} regions with no detected near-infrared cluster counterpart 
($25-39$\% of the detected sources) represent 
the youngest sites of star formation, with ages of up to approximately 
5\,Myr and mostly intermediate mass ($\simeq 10^5\,{\rm M}_{\odot}$)
ionizing clusters. For these two galaxies, and 
within the present detection threshold we can only 
detect coincidences ($4-10$\% of the detected sources) 
between an H\,{\sc ii} region and a near-infrared
star cluster  for the most massive 
star clusters ($\simeq 10^6\,{\rm M}_{\odot}$) during the first 7\,Myr
of their evolution. If there is significant extinction during 
the first million years, we may not 
detect the youngest star forming regions, and hence the observed
fractions of H\,{\sc ii} regions and coincidences will be lower limits. 
The identified near-infrared SSCs with no detectable 
Pa$\alpha$ emission represent the ``old'' population ($53-66$\% of 
the detected sources), with ages of between 7 and $20-40$\,Myr. 
Older clusters possibly created in this or 
previous episodes of star formation are likely to 
exist in these systems but cannot be identified
with the present detection threshold. 
Our study demonstrates 
that Pa$\alpha$ narrow-band imaging of  
LIRGs and interacting galaxies identifies 
the youngest sites of star formation 
that could be otherwise missed by near-infrared broad-band 
continuum surveys.

\end{abstract}


\section{Introduction}
Infrared galaxies (Rieke \& Low 1972), and in particular 
nearby luminous and ultraluminous infrared galaxies (LIRGs
and ULIRGs, with 
$L_{\rm IR} = 10^{11} - 10^{12}\,$L$_\odot$ and 
$L_{\rm IR} > 10^{12}\,$L$_\odot,$ respectively) have long been
recognized as one of the best laboratories to study the 
process of violent star formation in the Local Universe.
The dust-rich environments of LIRGs and ULIRGs are thought 
to be similar to the conditions in which
star formation 
occurred at high redshift (Ivison et al. 2000 and references therein).

It is clear now from both the theoretical 
and the observational viewpoints that interactions play a major
role in enhancing star formation 
throughout galaxies. In fact, 
the fraction of interactions/mergers increases 
from $20-30\%$ for $L_{\rm IR} < 10^{10}\,$L$_\odot$ up to
$70-95\%$ for $L_{\rm IR} > 10^{12}\,$L$_\odot$.
The process of merging with accompanying super-starbursts 
appears to be an important stage in galactic evolution, possibly even 
converting spiral galaxies into ellipticals (see Sanders \& Mirabel
1996 for a review). Moreover, it has been recently suggested that
most interacting/merging systems have gone or will go
through an ultraluminous phase, depending on how efficiently 
molecular gas can be funneled into the nuclear regions. This 
may imply that most of ULIRG class systems may spend a large portion
of their lifetime in the less powerful class of LIRGs 
(Murphy et al. 2001).

\begin{deluxetable}{cccllll}
\small
\tablewidth{18cm}
\tablecaption{Sample and observations.}
\tablehead{\colhead{Galaxy}  & \colhead{$\log L_{\rm IR}$} &
\colhead{Dist} & \colhead{Morphology}  
& \colhead{Filters} & \colhead{Linear} & \colhead{Program}\\
        &  \colhead{(L$_\odot$)} & \colhead{(Mpc)}& & &\colhead{Scale}}
\startdata
NGC~6808  &  10.94 & 46 & Isolated & NIC3 F160W, F190N & 45\,pc pixel$^{-1}$ 
& 7919\\
NGC~5653   & 11.01 & 47 & Isolated & NIC3 F160W, F190N & 45\,pc pixel$^{-1}$ 
& 7919\\
Zw~049.057 & 11.22 & 52 & Isolated, edge-on & NIC2 F160W, F187N, F190N&
19\,pc pixel$^{-1}$&7219 \\
NGC~3256   & 11.48 & 37 & Advanced merger & NIC2 F160W, F187N, F190N&
14\,pc pixel$^{-1}$ &7251 \\
NGC~1614   & 11.62 & 64 & Advanced merger & NIC2 F160W, F187N, F190N&
24\,pc pixel$^{-1}$ & 7218\\
VV~114     & 11.62 & 80 & Interacting pair & 
NIC2 F160W, NIC3 F187N, F190N & 29, 78\,pc pixel$^{-1}$ & 7219\\
NGC~6240  & 11.82 & 97 & Advanced merger & NIC2 F160W, F187N, F190N &
36\,pc pixel$^{-1}$ & 7219\\
Arp~299$^*$    & 11.91 & 42 & Interacting pair & NIC2 F160W, F187N, F190N&
15\,pc pixel$^{-1}$ & 7218\\ 
\enddata
\tablecomments{The infrared luminosity corresponds 
to the spectral range between 8 and $1000\,\mu$m.\\
$^*$The two components of Arp~299 are usually referred to as
IC~694 (the Eastern component) and NGC~3690 (the Western component). }
\end{deluxetable}

In last decade or so, mainly with {\it HST} observations, we have 
started to get the  first glimpses into the detailed 
nature (that is, on scales of a few tens of parsecs)
of the extreme star formation processes in LIRGs and interacting
galaxies.  The discovery,
both in the optical and the 
near-infrared, of a large population of massive (young) star clusters, 
the so-called super star clusters (SSCs), has shown that 
a significant fraction of the massive star formation is
occurring there   
(see among others, Whitmore et al. 1993; 
Schweizer et al. 1996;  
Zepf et al. 1999; Alonso-Herrero et al. 2000, and 2001a -- AAH00, AAH01;
Scoville et al. 2000 and references therein). 
This population of SSCs is not only inherent to LIRGs and 
ULIRGs, but it has also been identified in groups of galaxies 
(e.g., Gallagher et al. 2001) or even isolated barred or
ringed galaxies (e.g., 
Alonso-Herrero, Ryder, \& Knapen 2001b; 
Maoz et al. 2001 and references therein). For a recent review on the
subject of SSCs in external galaxies, we refer
the reader to Whitmore (2000).

While a lot of effort is being devoted to understanding the properties of
SSCs, {\it HST}/NICMOS imaging has  only recently 
revealed a population of bright H\,{\sc ii} regions in two LIRGs, 
Arp~299 and NGC~1614 (AAH00; AAH01). A large fraction of
these H\,{\sc ii} regions show luminosities in excess of that of 
30 Doradus, the prototypical giant H\,{\sc ii} region. 
One of the main difficulties in quantifying the age of the stellar populations
in LIRGs and interacting galaxies is breaking the 
age-extinction degeneracy. This usually translates into only rough 
age estimates for SSCs
(from photometric data)  ranging between 5 and  900\,Myr  
(see the recent review by Whitmore 2000). 
H\,{\sc ii} regions, on the other hand,  will highlight the youngest 
regions of star formation, with ages of $< 5-10\,$Myr, as 
these are the lifetimes of the O and B
stars required to ionize the gas. Clearly,
understanding the properties of H\,{\sc ii} regions and SSCs, 
and their relation  at high spatial resolution  will 
provide further insight into the nature of the star formation 
processes in LIRGs.

In this paper we present a study of the {\it detailed} 
(tens to a few hundred parsecs) properties of the  
star forming regions (H\,{\sc ii} regions and star clusters)
of a sample of 8 LIRGs. This paper is organized as follows. 
Section~(2) describes the observations, data reduction and the production 
of the H\,{\sc ii} region and star cluster catalogs. 
In Section~(3) we establish the overall morphology 
of star forming regions in LIRGs and its relation with the dynamical 
stage of the galaxy. In Section~(4) the statistical 
properties of H\,{\sc ii} regions 
in LIRGs are analyzed and compared with those of normal galaxies observed
at comparable spatial resolutions. 
The spatial 
distribution of H\,{\sc ii} regions and star clusters, their 
relative numbers and the age sequence are 
analyzed in Section~(5).  Our conclusions are presented 
in Section~(6).

\begin{figure*}
\figurenum{1}
\plotfiddle{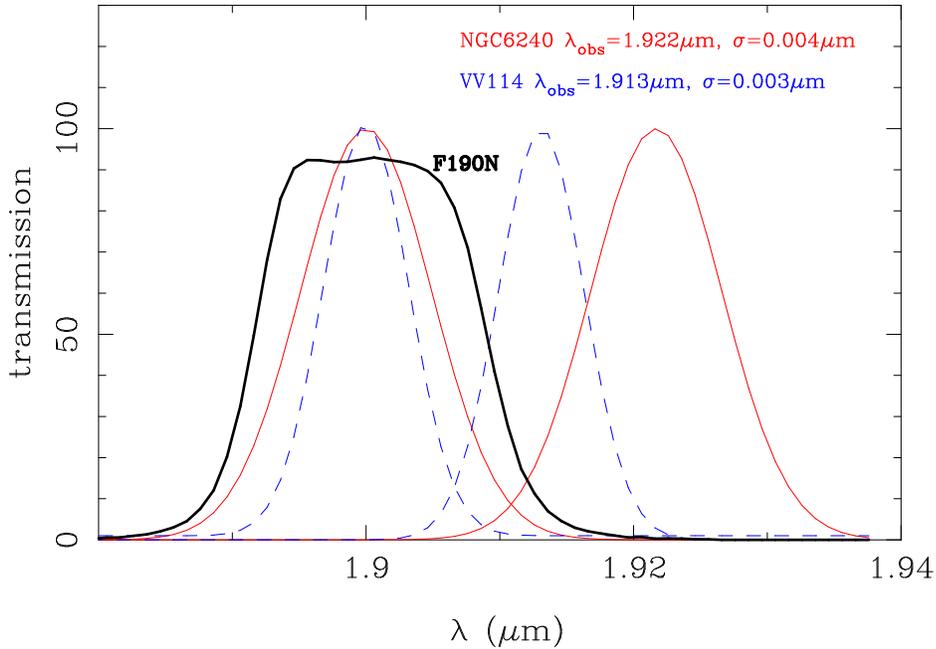}{425pt}{0}{70}{70}{-200}{150}
\vspace{-6cm}
\caption{Observed Pa$\alpha$ lines for VV~114 (dashed line) 
and NGC~6240 (solid line)
depicted as Gaussian functions with an arbitrary flux, and 
the observed values of the wavelength and line width.  
The same emission lines are also displayed as if 
their observed wavelengths corresponded to that of the 
center of the filter. We also show the transmission curve of the F190N 
filter as the solid thick line. }
\end{figure*}

\section{Observations}

The large amount of extinction routinely present in LIRGs, and in 
particular, the fact that active star forming regions 
are expected to contain non-negligible amounts of dust, prompted us to  
search the {\it HST} archive for infrared observations. The obvious
choice was  narrow-band 
Pa$\alpha$ and broad-band 
continuum imaging to identify H\,{\sc ii} regions and 
star clusters respectively. This resulted in a sample 
of eight LIRGs. The sample covers a range of infrared luminosities 
-- between $\log L_{\rm IR} =
10.94\,{\rm L}_\odot$ and $\log L_{\rm IR} =
11.82\,{\rm L}_\odot$ --  as well as 
a variety of dynamical stages: isolated galaxies, close pairs of
interacting galaxies and advanced mergers.
The sample is presented 
in Table~1\footnote{Note that although NGC~6808 is not technically a LIRG, 
its infrared luminosity ($\log L_{\rm IR} = 10.94\,{\rm L}_{\odot}$)
is close enough to the limit that it is included in our sample} 
in increasing order of infrared luminosity.

\subsection{Data reduction}

The observations of the LIRGS analyzed in this paper 
were obtained as part of a variety of {\it HST}/NICMOS GTO and GO 
programs, listed in Table~1. 
The Pa$\alpha$+continuum images were taken with the NIC2 and NIC3 cameras 
(pixel size 0.076\arcsec pixel$^{-1}$ and 0.2\arcsec \
pixel$^{-1}$, respectively) using the 
narrow-band filter F190N filter. At the distances of the LIRG sample, 
this filter ($\Delta \lambda/\lambda
\simeq 1\%$) contains the Pa$\alpha$ emission line  and the adjacent
continuum at $1.90\,\mu$m.  For the continuum subtraction, images
through the  F187N filter 
were used, except for NGC~5653 and NGC~6808 (from B\"oker et al. 1999 survey)
for which  broad-band F160W filter observations were employed instead. 
The field of view of the images is $19.5\arcsec \times 19.5\arcsec$
and $51.2\arcsec \times 51.2\arcsec$ for the NIC2 and NIC3 observations, 
respectively. All the continuum images used to identify the star 
clusters were
observed through the F160W filter which represents a good compromise
between the better spatial resolution at shorter wavelengths and lower
extinction at longer wavelengths.

In Table~1 for each galaxy we list the 
infrared luminosity, distance (assuming $H_0 = 75\,{\rm km\,s}^{-1}\,
{\rm Mpc}^{-1}$), large scale morphology, cameras and filters used,  
the corresponding linear scale in parsec per pixel and the {\it HST} 
program number of the observations.

The images were reduced using {\sc nicred} (McLeod et al. 1997) 
routines, which involve subtraction of the first readout, 
dark current subtraction on a readout-by-readout basis, correction
for linearity and cosmic ray rejection (using FULLFIT), and flat fielding. 
For the data reduced we used on-orbit dark and flatfield images.
The angular resolution
(FWHM) of the NIC2 and NIC3 images is approximately 
0.14\arcsec and 0.30\arcsec, respectively, as measured from the point
spread function (PSF) of stars in the images.  

\subsection{Flux calibration and correction for off-center emission 
lines}
The flux calibration of the F160W, F187N and F190N images was performed using
conversion factors based upon measurements of the standard star P330-E,
taken during the Servicing Mission Observatory Verification (SMOV)
program (M. J. Rieke, private communication 1999). 

The observed wavelength of Pa$\alpha$ 
for the two most distant galaxies in our sample, VV~114 and NGC~6240, 
is $1.9131\,\mu$m and $1.9217\,\mu$m respectively. Figure~1 shows 
the observed Pa$\alpha$ emission lines as Gaussian 
functions with line widths of $\sigma = 4.23 \pm 0.66 \times 10^{-3}\,\mu$m
for NGC~6240 (Goldader et al. 1997) and 
$\sigma = 3.2 \pm 0.3 \times 10^{-3}\,\mu$m for VV~114 (assuming 
that the line is unresolved and a 10\% error, Doyon et al. 1995, and 
Goldader et al. 1997). For comparison we also display 
the same emission lines as if they had been observed 
at the center of the F190N filter.  As 
can be seen from this figure for both galaxies, 
the observed Pa$\alpha$ emission lines are significantly displaced 
from the center of the narrow-band F190N filter. Assuming Gaussian 
line profiles, we estimate that the NICMOS filter may intercept 
only $\simeq 20\%$ and $\simeq 2\%$ of the full Pa$\alpha$ line
flux for these galaxies, respectively. Given the large and uncertain
correction implied to the true line flux, our results for these
two galaxies are only qualitative.

Throughout the paper we have converted the Pa$\alpha$ luminosity 
into the more commonly used H$\alpha$ luminosity, assuming case B 
recombination ($\frac{{\rm
H}\alpha}{{\rm Pa}\alpha} = 8.7$).

\subsection{Production of the H\,{\sc ii} region and 
star cluster catalogs}

The H\,{\sc ii} region catalogs were produced using the software {\sc
region}, kindly provided by Dr. C. H. Heller (see Pleuss, Heller, \& 
Fricke 2000,
and references therein for a detailed description).  {\sc region} is a
semi-automated method to locate and compute statistics of H\,{\sc ii}
regions in an image, based on contouring, taking into account the
local background. The lower limit for the size of an H\,{\sc ii} region
is set to 9 contiguous pixels, which for the distances of
the galaxies in our sample  corresponds to {\it minimum} linear
sizes (diameters) of between 42 and 234\,pc. 
Each pixel must have an intensity above
the local background of at
least three times the rms noise of that local background (see Rand 1992
for more details on the criteria employed). After
identifying the H\,{\sc ii} regions, the program measures their
position, size (area) and luminosity by subtracting the closest local
background from the observed flux. For more details on the 
method and a discussion on the spatial resolution effects see 
Pleuss et al. (2000), and Alonso-Herrero \& Knapen (2001).

In Section~5 we will compare the number of detected 
H\,{\sc ii} regions with the number of star clusters. To this 
end we need to determine the 
magnitude to which the $H$-band luminosity distributions 
of the star clusters are complete. The positions of the 
star clusters were identified from the $1.6\,\mu$m 
continuum images using the DAOFIND task in {\sc iraf}. 
As discussed in AAH00 for Arp~299, we find 
that the star clusters in most of the systems studied here are 
located on a local background with a substantial gradient, and a large 
degree of crowding. Thus
an automated method to perform photometry as that provided by DAOFIND 
may yield to systematic errors, especially 
for the faintest star clusters. 
For this reason we have 
performed aperture photometry on the clusters 
of the three galaxies with the best 
spatial resolution (NGC~3256 and the two components
of Arp~299) for which the results will 
be discussed in Section~5.

The photometry was obtained
through a relatively small aperture (${\rm diameter} =0.456\arcsec$), 
so an aperture 
correction was applied to account for the total flux. The absolute 
$H$-band magnitudes of the star clusters in these two systems 
range up to $M_H=-17\,$mag. The detection lower limit depends on the 
underlying galaxy background and the degree of
crowding. We find that the luminosity distribution of
the star clusters is complete down to 
$M_H=-14\,$mag and $M_H=-13.5\,$mag for NGC~3256 and Arp~299, respectively.  
This includes approximately $77\pm5\%$ of the detected 
clusters in both galaxies. The error accounts for uncertainties in 
the local background subtraction.
For the luminosities and properties of the 
brightest star clusters in NGC~1614 we refer the reader to 
AAH01. Scoville et al. (2000) presented photometry for  
the brightest near-infrared star clusters in Zw049.057, NGC~6240 
and VV~114. As is the case for NGC~3256 and Arp~299, the 
cluster $H$-band magnitudes are found to range up to $M_H \simeq 
-17.2\,$mag.

\begin{deluxetable}{ccccc}
\small
\tablewidth{12cm}
\tablecaption{Line emission and H\,{\sc ii} region properties in LIRGs.}
\tablehead{\colhead{Galaxy}  & \colhead{FOV of Pa$\alpha$ } &
\colhead{$\log L({\rm H}\alpha)_{\rm tot}$} & 
\colhead{$L_{\rm nuc}/L_{\rm tot}$}  
& \colhead{$\log \bigl(L_{\rm IR}/L({\rm H}\alpha)_{\rm tot}\bigr)$}\\
 & &\colhead{(erg s$^{-1}$)} & & }
\startdata
NGC~6808  &  $11\,{\rm kpc}\times 11\,{\rm kpc}$ & 41.38 & $\simeq 0$ & 3.1\\
NGC~5653  &  $11\,{\rm kpc}\times 11\,{\rm kpc}$ & 41.72 & $\simeq 0$ & 2.9\\
Zw~049.057 & $4.9\,{\rm kpc}\times 4.9\,{\rm kpc}$ & 41.21 & -- & 3.6\\
NGC~3256   & $3.5\,{\rm kpc}\times 3.5\,{\rm kpc}$ & 42.11 & 0.41 & 3.0 \\
NGC~1614   & $6.4\,{\rm kpc}\times 6.4\,{\rm kpc}$ & 42.60 & 0.65 & 2.5\\
VV~114     & $9.3\,{\rm kpc}\times 9.3\,{\rm kpc}$ & 42.55$^*$ & 0.37 & 
2.7$^*$\\
IC~694     & $3.8\,{\rm kpc}\times 3.8\,{\rm kpc}$ & 41.95 & 0.58 & 2.6\\ 
NGC~3690   & $3.8\,{\rm kpc}\times 3.8\,{\rm kpc}$ & 42.16 & 0.18 & 3.1\\ 
NGC~6240   & $9.2\,{\rm kpc}\times 9.2\,{\rm kpc}$ & 43.19$^*$ 
& $\simeq 1^*$& 2.2$^*$\\
\enddata
\tablecomments{``FOV of Pa$\alpha$'' is  the area 
imaged through this emission line.\\
$^*$Because of the large correction needed to account for the 
Pa$\alpha$ emission line flux missing in the F190N filter (Section~2.2), these 
values are uncertain.}
\end{deluxetable}

\begin{figure*}
\figurenum{2a}
%\plotfiddle{figure2a.ps}{425pt}{0}{90}{90}{-280}{-250}
%\vspace{6.5cm}
\caption{{\it Bottom left panel:} Grey-scale continuum image at $1.6\,\mu$m (on a 
logarithmic scale) of
NGC~6808. {\it Bottom right  panel:} Grey-scale continuum subtracted Pa$\alpha$ image 
(on a  logarithmic scale). {\it Upper panel:} The contours (on a linear scale) 
are the continuum-subtracted 
Pa$\alpha$ emission. The stars (blue stars in the on-line color version 
of Figure~2) are the positions of the 
identified near-infrared
star clusters, whereas the open circles (red circles in the 
on-line color version of this figure) are the emission peaks of
the identified H\,{\sc ii} regions. For reference, we mark the position of the 
nucleus with an 'N'.}
\end{figure*}

\begin{figure*}
\figurenum{2b}
%\plotfiddle{figure2b.ps}{425pt}{0}{90}{90}{-280}{-250}
\caption{{\it Continued:} Same as Figure~2a but for NGC~5653.}
\end{figure*}

\begin{figure*}
\figurenum{2c}
%\plotfiddle{figure2c.ps}{425pt}{-90}{70}{70}{-280}{550}
%\vspace{-6cm}
\caption{{\it Continued:} The left panel is the continuum image of Zw~049.057. 
The right panel are the contours of the Pa$\alpha$ emission, with only the positions of 
star clusters shown.}
\end{figure*}

\begin{figure*}
\figurenum{2d}
\plotfiddle{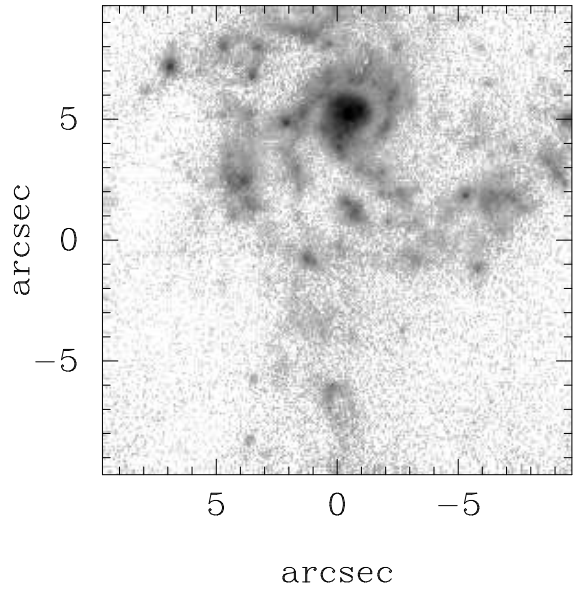}{425pt}{0}{90}{90}{-280}{-250}
\vspace{6.5cm}
\caption{{\it Continued:} Same as Figure~2a but for NGC~3256.}
\end{figure*}

\begin{figure*}
\figurenum{2e}
%\plotfiddle{figure2e.ps}{425pt}{0}{90}{90}{-280}{-250}
%\vspace{6.5cm}
\caption{{\it Continued:} Same as Figure~2a but for NGC~1614.}
\end{figure*}

\begin{figure*}
\figurenum{2f}
%\plotfiddle{figure2f.ps}{425pt}{0}{90}{90}{-280}{-250}
%\vspace{6.5cm}
\caption{{\it Continued:} Same as Figure~2a but for VV~114.
The NIC2 F160W continuum image was degraded to the 
resolution of the NIC3 F190N image before the identification
of the star clusters (as shown here).
The nuclei of the system are marked with the symbols 'NE', 'NW' and 'W'
(following Soifer et al. 2001 notation).}
\end{figure*}

\begin{figure*}
\figurenum{2g}
%\plotfiddle{figure2g.ps}{425pt}{0}{90}{90}{-280}{-250}
%\vspace{6.5cm}
\caption{{\it Continued:} Same as Figure~2a but for IC~694, 
the Eastern component of Arp~299.}
\end{figure*}

\begin{figure*}
\figurenum{2h}
%\plotfiddle{figure2h.ps}{425pt}{0}{90}{90}{-280}{-250}
%\vspace{6.5cm}
\caption{{\it Continued:} Same as Figure~2a but
for NGC~3690, the Western component of Arp~299.}
\end{figure*}

\begin{figure*}
\figurenum{2i}
%\plotfiddle{figure2i.ps}{425pt}{-90}{70}{70}{-280}{500}
%\vspace{-4cm}
\caption{{\it Continued:} The left panel in the continuum image of NGC~6240.
The right panel are the contours of the Pa$\alpha$ emission with the 
positions of the clusters and H\,{\sc ii} regions marked as in 
Figure~2a.}
\end{figure*}

\section{Morphology of the Pa$\alpha$ line emission in LIRGs}

The top panels of Figure~2 are the 
contours of the continuum-subtracted Pa$\alpha$ line emission
 for the galaxies in our sample. We show the locations of 
the near-infrared star  clusters and the emission  peaks 
of the H\,{\sc ii} regions as stars and open circles
respectively (see  Section~5). The bottom left panels of 
Figure~2 are the continuum emission images at $1.6\,\mu$m, 
whereas the bottom right panels are the continuum 
subtracted Pa$\alpha$ images for the same  
fields of view. The images are displayed with the 
original {\it HST} orientation. These figures illustrate the 
variety of Pa$\alpha$
line emission morphologies found in LIRGs. 

The isolated galaxies in our sample (NGC~6808, NGC~5653 and 
Zw~049.056) exhibit H\,{\sc ii} regions along the spiral 
arms. NGC~6808, the LIRG in our sample with the lowest infrared 
luminosity shows the typical H\,{\sc ii} region distribution of 
a spiral galaxy. 
Zw~049.056 appears as a  highly inclined galaxy, and 
only shows faint Pa$\alpha$ emission 
along the disk of the galaxy, perhaps as a result of 
the high extinction there (Scoville et al. 2000). NGC~5653 
shows asymmetric Pa$\alpha$ emission, with most and the brightest
H\,{\sc ii} regions located  
in one of the spiral arms. This galaxy has been 
classified as a lopsided galaxy (Rudnick, Rix, \& Kennicutt 2000).
Lopsidedness is usually assumed to be a dynamical indicator of a  
weak tidal interaction, that is, when the interaction does not
destroy the disk of the galaxy. No bright nuclear emission is 
detected in  NGC~5653 and NGC~6808.

The Pa$\alpha$ line emission of 
interacting/merging systems shows a more complex morphology, differing
significantly from the stellar light distribution (continuum images). 
A common denominator for all these systems is the bright emission from the 
nuclei of the galaxies. The Pa$\alpha$ nuclear emission is 
usually compact, extending over scales of  
$500-650\,$pc, except in
NGC~1614 where the Pa$\alpha$ emission  in the form of 
a ring of giant H\,{\sc ii} regions  with 
an approximate diameter of 600\,pc (see AAH01 for more details).

In the interacting/merging systems, besides the bright nuclear 
Pa$\alpha$ emission, the 
recent star formation is also widely spread throughout the galaxies.
In NGC~3256 and NGC~1614 the Pa$\alpha$ emission extends along the spiral 
arms over at least 
4\,kpc and 6\,kpc (that is, the field of view of the {\it HST}/NICMOS images
for these galaxies). In the interacting systems Arp~299 and VV~114, 
H\,{\sc ii} regions are detected throughout the galaxies, and
at the interface region. This is
in good agreement with findings for ULIRGs in Murphy et al. (2001). 
The Pa$\alpha$ morphology in the 
merging/interacting systems clearly shows that 
the effects of extreme star formation can propagate throughout the 
galaxies and not only in the nuclei. Recently Bekki \&
Couch (2001) have run stellar and gas numerical simulations of 
galaxy mergers to find the most promising sites for formation of 
SSCs. They conclude that these SSCs are not only formed in the central 
regions, but also between the two interacting galaxies. 

NGC~6240 only shows Pa$\alpha$ emission 
originating in both nuclei (separated by 
only 0.8\,kpc). This is consistent with recent 
{\it HST}/WFPC2 H$\alpha$ imaging (Gerssen et al. 2001). The 
larger scale filamentary diffuse morphology in NGC~6240 has not been 
detected by the NICMOS observations, probably as a result of the 
Pa$\alpha$ line falling at the edge of the narrow-band filter 
(see Section~2.2). 

\begin{figure*}
\figurenum{3}
\plotfiddle{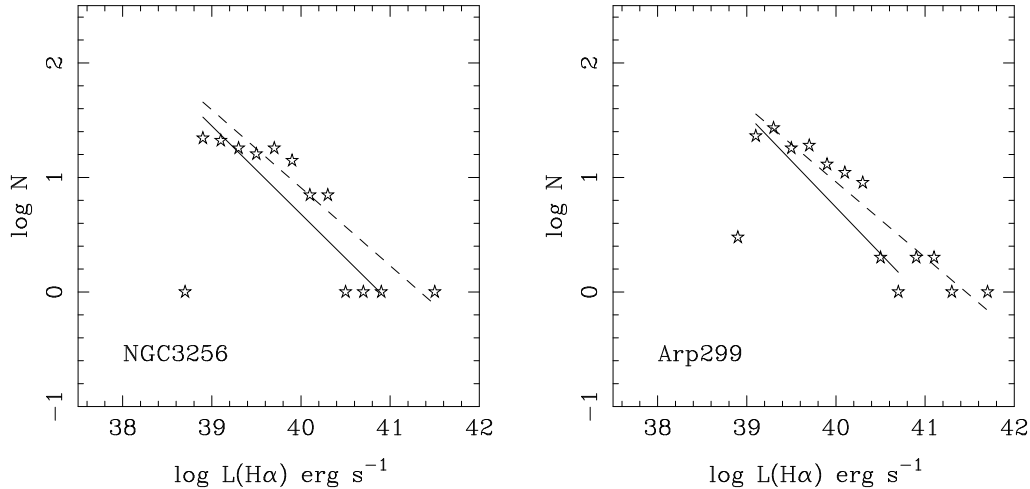}{425pt}{-90}{70}{70}{-240}{630}
\vspace{-8.cm}
\caption{H\,{\sc ii} region LFs for 
NGC~3256 and Arp~299. The solid lines are the power law fits excluding the 
nuclei of the galaxies, and the regions with strong 
star formation activity in Arp~299 (C and C'). For 
comparison the dashed lines represent the fits including 
all the sources.}
\end{figure*} 

As most of the infrared luminosity in LIRGs is believed 
to be due to dust heated in regions of strong star formation, the 
mid-infrared emission in LIRGs would be expected to 
trace the  Pa$\alpha$ emission. 
Recently Soifer et al. (2001) have obtained high resolution 
mid-infrared observations 
of LIRGs and reported that a substantial fraction ($>50\,\%$) of the 
infrared luminosity is generated in regions of sizes ranging from 100\,pc 
to a 1\,kpc, with the nuclear starbursts often dominating the 
mid-infrared luminosity. They also concluded that the $12\,\mu$m 
emission traces the current sites of {\it dusty} luminous 
star formation. This can 
be confirmed by the good correspondence between the mid-infrared and the 
Pa$\alpha$ emission morphologies of the galaxies in common with our 
study: Arp~299, VV~114 and NGC~1614. 

\begin{deluxetable}{ccccccc}
\small
\tablewidth{16cm}
\tablecaption{Statistical properties of H\,{\sc ii} regions.}
\tablehead{\colhead{Galaxy}  & \colhead{$\log L({\rm H}\alpha)_{\rm br}$} & 
\colhead{$\log <L({\rm H}\alpha)>_3$} & 
\colhead{$\log L({\rm H}\alpha)_{\rm median}$} & \colhead{${\rm size}_{\rm large}$} &  
\colhead{$<{\rm size}>_3$} & \colhead{${\rm size}_{\rm median}$}\\
&\colhead{(erg s$^{-1}$)} & \colhead{(erg s$^{-1}$)} & \colhead{(erg s$^{-1}$)}& (pc) & (pc) & (pc) }
\startdata
\multicolumn{7}{c}{LIRGs}\\
\hline
NGC~3256  & 41.53 & 41.24 & 39.51 & 308 & 253 & 80\\
IC~694    & 41.71 & 41.29 & 39.56 & 448 & 298 & 82\\ 
NGC~3690  & 41.43 & 41.33 & 39.72 & 346 & 320 & 94\\ 
NGC~1614  & 41.01 & 41.00 & 39.89$^a$ & 210$^a$ & 205$^a$ & 148$^a$ \\
NGC~6808  & 40.18 & 40.11 & 39.22 & 461 & 423 & 201\\
NGC~5653  & 41.02 & 40.74 & 39.49 & 551 & 468 & 199\\
VV~114    & 41.94$^b$ & 41.79$^b$ & 40.52$^b$ & 920 & 842 & 382 \\
\hline
\multicolumn{7}{c}{Comparison Normal Galaxies}\\
\hline
NGC~4192  & 39.52 & 39.39 & 38.52  & 158 & 147 & 76 \\
NGC~4389  & 39.44 & 39.36 & 38.70  & 151 & 134 & 74\\
IC~750    & 39.89 & 39.73 & 38.62  & 191 & 184 & 72\\
NGC~4102  & 40.48 & 40.32 & 38.32  & 263 & 223 & 80\\ 
\enddata
\tablecomments{$^a$ Statistics excluded the bright H\,{\sc ii}
regions in the ring of star formation, as blending effects are present.\\
$^b$ Uncertain because of the large correction needed to account for the 
Pa$\alpha$ emission line flux missing in the F190N filter (Section~2.2).}
\end{deluxetable}

In Table~2 we list the total H$\alpha$ luminosity from 
the detected H\,{\sc ii} regions, along with 
the percentage of the Pa$\alpha$ emission 
originating in the nuclei of galaxies. These fractions should be considered
only as lower limits as the compact nuclear regions are usually 
more reddened than the regions where the more diffuse H\,{\sc ii} regions 
are located (see e.g., AAH00; L\'{\i}pari et al. 2000; Scoville et al. 2000; 
see also Murphy et al. 2001 for ULIRGs).
Nevertheless it is apparent that the 
nuclear to total emission ratios vary significantly from system to system, 
and show similarities with the same ratios computed for the infrared 
luminosities (Soifer et al. 2001).
In the isolated systems most of the star formation is 
occurring in the disks of the galaxies, whereas in the interacting/merging 
systems the location of the 
star forming regions depends on factors such as  
the age of the interaction process and 
the physical properties of the interacting galaxies (among others, 
the initial gas content and mass ratios of the interacting galaxies, 
see e.g., Bekki \& Couch 2001). Table~2 also gives  
the ratios between the infrared luminosity and the total 
H$\alpha$ luminosity for the LIRGs in our sample. These ratios 
are found to be similar for all the galaxies except for Zw~049.057 
where the Pa$\alpha$ emission is probably affected by high extinction, and 
NGC~6240 for which the Pa$\alpha$ luminosities are uncertain
as discussed in Section~2.2.

\section{Statistical properties of giant H\,{\sc ii} regions in LIRGs} 

\subsection{Luminosities, Sizes and Luminosity Functions}

We have measured the  luminosity and diameter of the identified 
H\,{\sc ii} regions in our sample of LIRGs.  For 
each galaxy we determined the  H$\alpha$ luminosity for the 
brightest H\,{\sc ii} region, for the  three brightest H\,{\sc ii} 
regions (first-ranked H\,{\sc ii} regions) and the median H\,{\sc ii}
region, as well as the diameter for the largest H\,{\sc ii} region 
(which is not necessarily the brightest H\,{\sc ii} region), 
for the first-ranked H\,{\sc ii} regions and the median H\,{\sc ii}
regions. The results are summarized 
in the first part of Table~3, where galaxies are sorted in order 
of decreasing spatial resolution. 
As discussed in  Alonso-Herrero \&
Knapen (2001) and others, even at {\it HST} spatial resolutions the problem 
of defining an H\,{\sc ii} region remains. For instance, 
at increasing higher spatial resolutions different emission peaks within 
an H\,{\sc ii} region may be 
identified as individual H\,{\sc ii} regions thus cataloging them as
smaller and less luminous H\,{\sc ii} regions. Conversely, 
as the spatial resolution becomes poorer,  blending effects will 
become more frequent. This effect is clearly seen in the increasing
median sizes of H\,{\sc ii} regions in the LIRG sample for decreasing 
spatial resolutions (Table~3). 

All the LIRGs in our sample contain a significant number of exceptionally 
bright H\,{\sc ii} regions (i.e., the median 
values of the H\,{\sc ii} region luminosity distribution) 
with H$\alpha$ luminosities comparable to that of the giant H\,{\sc ii}
region 30 Doradus ($\log L({\rm H}\alpha) \simeq 39.70\,{\rm erg\,s}
^{-1}$). The first-ranked H\,{\sc ii} regions in the 
interacting/merging galaxies -- usually coincident with the nuclei of
the galaxies -- are between  one and two orders of magnitude brighter than 
the median H\,{\sc ii} regions. As can be seen from Table~3  the 
luminosities of the brightest and first-ranked H\,{\sc ii} of the two 
isolated galaxies, even at poorer spatial resolutions, 
are markedly less luminous than their counterparts in interacting/merging
systems. Unfortunately the small number of LIRGs in our sample does not
allow us to establish whether isolated LIRGs tend to have 
fainter H\,{\sc ii} regions than interacting LIRGs.

The statistical properties of 
H\,{\sc ii} regions are  represented with LFs, 
usually fitted by a power law:

\begin{equation}
{\rm d}N(L) = A\,L^\alpha\,{\rm d}L
\end{equation}

\noindent where $\alpha$ is the index of the power law
with values of $\alpha =-2.0\pm0.5$ for extragalactic 
H\,{\sc ii} regions (Kennicutt, Edgar, \& Hodge 1989). We have constructed 
H\,{\sc ii} region LFs for the two LIRGs in 
our sample with a sufficiently high number of detected H\,{\sc ii}
regions: Arp~299 and NGC~3256 (shown in Figure~3). We have fitted 
the slopes of the LFs excluding the nuclei and 
the regions of strong star formation activity in Arp~299 (as 
they may be blends of a few regions) and 
found values of $\alpha = -1.77\pm 0.10$ and 
$\alpha = -1.81\pm 0.16$ for NGC~3256 and Arp~299, respectively.
If we include all H\,{\sc ii} regions in the fit, the 
values of the slopes are $\alpha = -1.68\pm 0.08$ 
and $\alpha = -1.66\pm 0.05$ respectively. These slopes are 
within the values fitted for H\,{\sc ii} regions in the disks 
and circumnuclear regions of normal galaxies (e.g., Kennicutt et al. 
1989), and those found from fitting the luminosity functions
of star clusters (e.g., Zepf et al. 1999; Whitmore 2000).

\begin{figure*}
\figurenum{4}
\plotfiddle{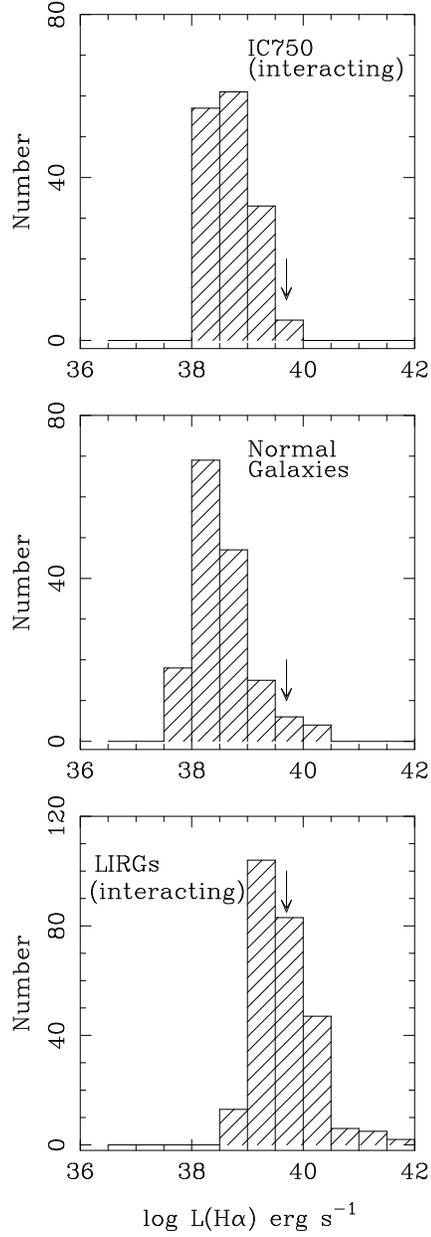}{425pt}{0}{70}{70}{-150}{-30}
\vspace{0.1cm}
\caption{Histograms comparing the H$\alpha$ luminosities 
of the 
H\,{\sc ii} regions detected in two LIRGs (Arp~299 and NGC~3256) 
in the bottom panel, 
normal galaxies (NGC~4192, NGC~4389 and NGC~4102) in the 
middle panel and the interacting galaxy IC~750, in the top panel. 
All these galaxies
have been observed with comparable spatial resolutions.
The arrow indicates
the H$\alpha$ luminosity of the giant H\,{\sc ii} region 30 Doradus.}
\end{figure*}

\subsection{Comparison with normal galaxies}

A number of studies have analyzed the statistical 
properties of H\,{\sc ii}
regions (luminosities, sizes, density numbers) 
in the disks of normal spiral galaxies in terms of 
the morphological type of the galaxy and the arm-interarm regions 
(see for instance the classical
study of Kennicutt et al. 1989). Because of the effect of the 
spatial resolution on the properties of H\,{\sc ii} regions, we
chose to compare 
the statistical properties of the H\,{\sc ii} regions in LIRGs with 
those of the circumnuclear H\,{\sc ii} regions of normal galaxies 
(as derived from {\it HST}/NICMOS Pa$\alpha$ observations) studied by  
Alonso-Herrero \& Knapen (2001).  From this work, we have 
selected galaxies at the distance of the Virgo Cluster (${\rm Dist} \simeq 
17\,$Mpc), for which the 
NIC3 observations provide a 
resolution of 16\,pc pixel$^{-1}$, similar to that of 
NGC~3256, NGC~3690 and IC~694.

In the second part of Table~3 we summarize 
the statistical 
properties of H\,{\sc ii} regions of the  normal 
galaxies with the highest degree of central star forming activity 
(as measured from the central kpc H$\alpha$ luminosity): 
NGC~4192, NGC~4389, IC~750, and NGC~4102.

Figure~4 shows histograms comparing 
the H$\alpha$ luminosities of  H\,{\sc ii} regions in LIRGs and normal 
galaxies. We have divided the normal galaxies into 
two groups, isolated  galaxies (NGC~4192, NGC~4389 and 
NGC~4102) and 
the interacting galaxy IC~750.
In this figure we also indicate the H$\alpha$ luminosity 
of 30 Doradus, the prototypical giant H\,{\sc ii} region. 
These  histograms not only show that giant H\,{\sc ii} regions are more common 
in LIRGs than in normal galaxies -- even when comparing with the interacting 
system IC~750 -- but also that the median H\,{\sc ii} regions
in LIRGs are at least an order of magnitude brighter than in normal 
galaxies.

The large population of giant  H\,{\sc ii}
regions in LIRGs -- with a significant fraction being 
more luminous than  30 Doradus -- 
is unprecedented in normal galaxies. 
There are a number of possible explanations. We may just be  
seeing the extended tail of the H\,{\sc ii} region LF which translates 
into a larger number of the more 
massive ionizing star clusters, a extremely young population 
of star forming regions in LIRGs, or   
perhaps the bright H\,{\sc ii} regions in LIRGs may represent
aggregations of normal H\,{\sc ii} regions. The typical (median) 
diameters for H\,{\sc ii} regions in normal galaxies and LIRGs observed
at the same spatial resolution 
are comparable (see Table~3). This seems to argue against the 
last possibility. Even though the first ranked H\,{\sc ii}
regions in LIRGs are typically a factor of two larger than in normal 
galaxies, the luminosities of the first-ranked (as well as median)
H\,{\sc ii} regions in LIRGs  would require 
aggregations of ten ``normal'' H\,{\sc ii} regions. 
Other possibilities are then evolutionary effects and 
the upper limit to the mass of the ionizing clusters.

As can be seen from the simulations of H\,{\sc ii}
region LFs by Oey \& Clark (1998) the age of the 
stellar population is one of the factors determining 
the high luminosity end of the H\,{\sc ii} region LF. 
For a population of H\,{\sc ii} regions formed in a 
single burst, the high luminosity end of the LF as well
as the slope vary as the population ages. In the case of 
continuous creation of H\,{\sc ii} regions the high luminosity 
end of the LF remains approximately constant. The youth of a population 
of coeval H\,{\sc ii} regions could be a possibility 
to explain the excess of bright H\,{\sc ii} regions in LIRGs. As we shall 
see in the 
next section however, the H\,{\sc ii} regions in LIRGs show a range of 
ages, which means that the extreme luminosities of the H\,{\sc ii}
regions in LIRGs are not due to the fact that they are all very young. 

Since the fitted slopes of the H\,{\sc ii} region LFs 
of two LIRGs are consistent
with the values found in normal galaxies, the extraordinary luminosities 
of H\,{\sc ii} regions in LIRGs are simply reflecting a larger number of  
massive star clusters in LIRGs, rather than an anomalous stellar mass 
distribution. As discussed in Bekki \& Couch (2001), the 
physical conditions (high gas pressure and density of the 
interstellar medium) necessary for the formation of super star clusters 
can only be achieved in systems undergoing a rapid transfer of gas to the 
central regions, that is, interacting galaxies. It is unlikely 
that these conditions are met in isolated disk galaxies.

\section{Giant H\,{\sc ii} regions and their relation to SSCs}

\subsection{Spatial distribution of the H\,{\sc ii} regions and 
near-infrared star clusters}
Figure~2 shows that the emission peaks of 
the H\,{\sc ii} regions are generally 
not spatially coincident with the location of the 
near-infrared star clusters. To quantify this, 
we have cross-correlated the positions 
of the H\,{\sc ii} regions and the near-infrared 
clusters. We counted a coincidence
when the spatial separation between an H\,{\sc ii} 
region and a cluster was equal or less than 
0.15\arcsec--0.11\,\arcsec \ (for the NIC2 images) and 
0.40\arcsec--0.30\,\arcsec (for the NIC3 images), that is, 
a separation of 2 and 1.5 pixels respectively. 

The results from the cross-correlation 
are presented in the first part of Table~4 where we list the 
number of detected H\,{\sc ii} regions, 
and star clusters, together with the number of coincidences.
As was clear from Figure~2, the number of coincidences is relatively 
small. The increasing number 
of coincidences (intermediate age H\,{\sc ii} regions/clusters)
for NGC~5653, NGC~6808 and VV~114 may be the result of the poorer spatial 
resolutions of the NIC3 images, which implies that a coincidence 
is counted when 
the spatial separations are $68-117\,$pc ($90-156\,$pc 
for the 2 pixel separation). For comparison, the spatial separations 
for the galaxies observed with the NIC2 camera are $21-38\,$pc 
(or $28-50\,$pc if we use the 2 pixel separation).
This lack of spatial coincidence between 
H\,{\sc ii} regions and clusters has also been found in 
barred and ringed galaxies with 
strong star formation activity (Alonso-Herrero
et al. 2001b; Maoz et al. 2001).

Theoretical arguments may offer an explanation
for the apparent offsets between the location of 
the SSCs and the H\,{\sc ii} regions in LIRGs. 
For instance, Tan \& McKee (2000) have proposed a model in which 
young clusters within clouds with 
masses $5 \times 10^4 - 5 \times 10^5\,
{\rm M}_\odot$ can survive for up to 3\,Myr  and 
allow for the formation of gravitationally bound clusters 
(which in turn may evolve into globular clusters)
in the presence of vigorous feedback (Wolf Rayet
winds and supernovae). According to their model, after 
approximately 3\,Myr,  the gas will be  
disrupted and the star cluster will emerge from its cocoon, and will 
be visible. The youngest 
SSCs are still embedded in the (probably dusty) H\,{\sc ii} regions, 
and demonstrates that giant H\,{\sc ii} regions offer us an unprecedented view 
to the truly youngest (high-mass) star formation
activity. The SSCs on the other hand show a broader
range of ages as we shall  show in the next section.

There are also Galactic  examples of this situation 
in obscured giant H\,{\sc ii} regions, where a variety of 
complex morphologies including nebular emission, regions 
of high and variable extinction and centrally concentrated 
star clusters are seen (see Blum et al. 2001 for 
a census, and references therein). In particular these authors 
have established  an evolutionary sequence for 
their sample of giant H\,{\sc ii} regions (see their figure~1), 
based on the morphology and spectroscopic ages of the stars. In 
the youngest H\,{\sc ii} regions the cluster stars 
are still veiled by the hot dust from their birth cocoons; 
as the H\,{\sc ii} region ages, the  clusters become 
more revealed as the gas will be dispersed by energetic 
winds and radiation pressure from the hot stars, until 
finally most of the gas will have dispersed.

The lack of spatial correspondence between 
the near-infrared SSCs and the bright H\,{\sc ii} regions in LIRGs
is suggestive of a similar evolutionary sequence  as that seen in 
obscured Galactic giant H\,{\sc ii} regions. To first order, ignoring
possible effects of extinction and the observational detection limits, 
those H\,{\sc ii} regions 
with no near-infrared cluster counterpart 
will tend to be the youngest examples of star forming regions 
in which the gas is being ionized by the most
massive stars, while the star clusters will
still be embedded; we denote them as ``young H\,{\sc ii} regions/clusters''.
They account for between 20\% and 40\% of all the detected 
sources (star clusters and H\,{\sc ii} regions) in LIRGs
(see Table~4). The coincidences correspond to the case of an 
evolved H\,{\sc ii} region in which the near-infrared cluster is becoming 
``visible'', that is, the time period in 
which the cluster has begun to emerge from the natal dust. At the same time
the first red supergiants will start appearing so the cluster will
become bright in the near-infrared.  
The ``intermediate H\,{\sc ii} regions/clusters'' account for 
up to $30\%$ of the cluster + H\,{\sc ii} region population in LIRGs. 
We also find that the number of coincidences tends to increase for the 
more  luminous H\,{\sc ii} regions. For
 instance, in NGC~3256 we find that $\simeq 55\%$ of the high luminosity
H\,{\sc ii} regions have a new-infrared cluster counterpart (see 
next section).
And finally, the near-infrared star 
clusters with no H\,{\sc ii} region counterpart represent the 
case where the most massive 
(ionizing) stars have disappeared so there will
be no detectable gas emission; these will be referred to as  
``old clusters''. We estimate that they contribute between 56\% and 
72\% of the detected cluster population of LIRGs.

\subsection{Modelling of the H\,{\sc ii} region to near-infrared 
star cluster age sequence}

The age/coincidence argument can be developed further by using 
outputs of evolutionary synthesis models and the 
relative numbers of H\,{\sc ii} regions, coincidences and 
near-infrared star clusters. Obviously for a given galaxy, the number of
identified sources depends on the detection threshold of
the current observations. It will be easier to present this argument using the 
three galaxies with the best spatial resolutions: 
NGC~3256, IC~694 and NGC~3690. The argument however will  
be valid for any galaxy by 
changing the detection limits.

In Section~2.3 we established that the $H$-band luminosity 
functions of the star  clusters in NGC~3256 and Arp~299 are complete down to 
absolute magnitudes of $M_H = -14\,$mag and
$M_H=-13.5\,$mag, respectively. 
The H\,{\sc ii} region luminosity 
functions are approximately complete down 
to $\log L({\rm H}\alpha) =  
38.8-39\,{\rm erg\,s}^{-1}$ (previous section), which is equivalent to 
a number of ionizing photons of $\log N_{\rm Ly} = 50.7-50.9\,{\rm s}^{-1}$.
We can use these detection limits to compute the fractions 
of young, intermediate and old populations for the complete 
distributions of NGC~3256, IC~694 and NGC~3690 (second part of 
Table~4).  Using the complete distributions increases the H\,{\sc ii}
region fraction by $\simeq 5\%$, whereas the old  cluster fraction 
drops by 
$\simeq 5\%$, and  the coincidence fraction remains approximately constant.
We will make use of evolutionary synthesis models to 
establish  an age sequence, as well
as to reproduce 
the fractions of young, intermediate and old  clusters in these
systems. Our choice for the evolutionary synthesis models will be Starburst99 
(Leitherer et al. 1999) and those of Rieke et al. (1993). 

\begin{figure*}
\figurenum{5}
\plotfiddle{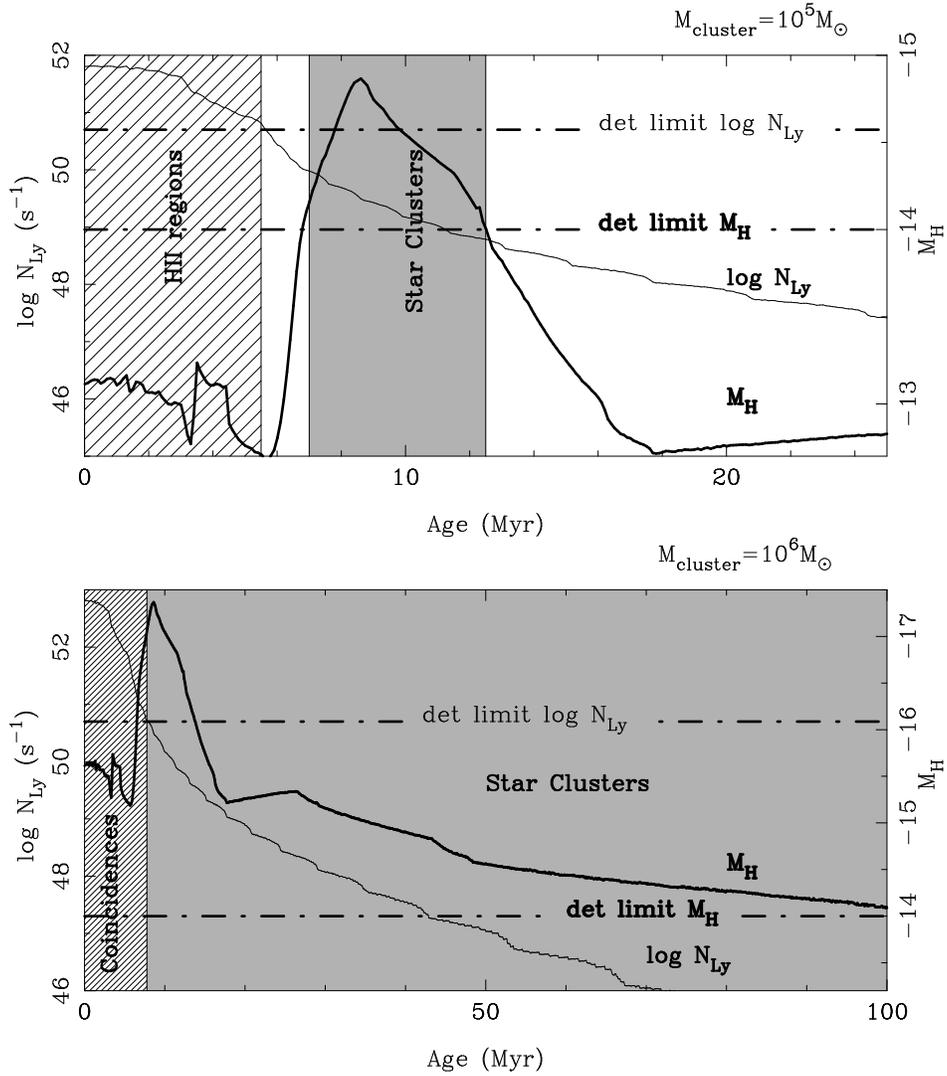}{425pt}{0}{70}{70}{-220}{30}
\vspace{-2.cm}
\caption{The solid lines are outputs of Starburst99 (Leitherer et al. 
1999) showing the time evolution of the number of ionizing photons 
($\log N_{\rm Ly}$, thin solid line, scale on the left hand side) 
and absolute $H$-band magnitude ($M_H$,
thick solid line, scale on the right hand side) 
for a $10^6\,{\rm M}_\odot$ cluster (bottom panel) 
and a $10^5\,{\rm M}_\odot$ cluster (top panel), for 
an instantaneous burst, a Salpeter IMF 
(between 1 and $100\,{\rm M}_\odot$) and solar metallicity.
The dashed-dotted lines represent the approximate 
detection thresholds for NGC~3256: $M_H=-14\,$mag and $\log N_{\rm Ly}
=50.7\,{\rm s}^{-1}$. An H\,{\sc ii} region with no cluster counterpart will 
be the case when $\log N_{\rm Ly}$ is above the detection threshold, but 
$M_H$ is not detectable yet (lightly hatched 
area). There will be a coincidence between 
an H\,{\sc ii} region and a near-infrared star cluster when 
both $N_{\rm Ly}$ and $M_H$ are above their 
respectively detection thresholds (closely hatched 
area). We will only observe a star cluster when $\log N_{\rm Ly} < 
50.7\,{\rm s}^{-1}$, but $M_H$ is still observable (grey area).}
\end{figure*}

\begin{figure*}
\figurenum{6}
\plotfiddle{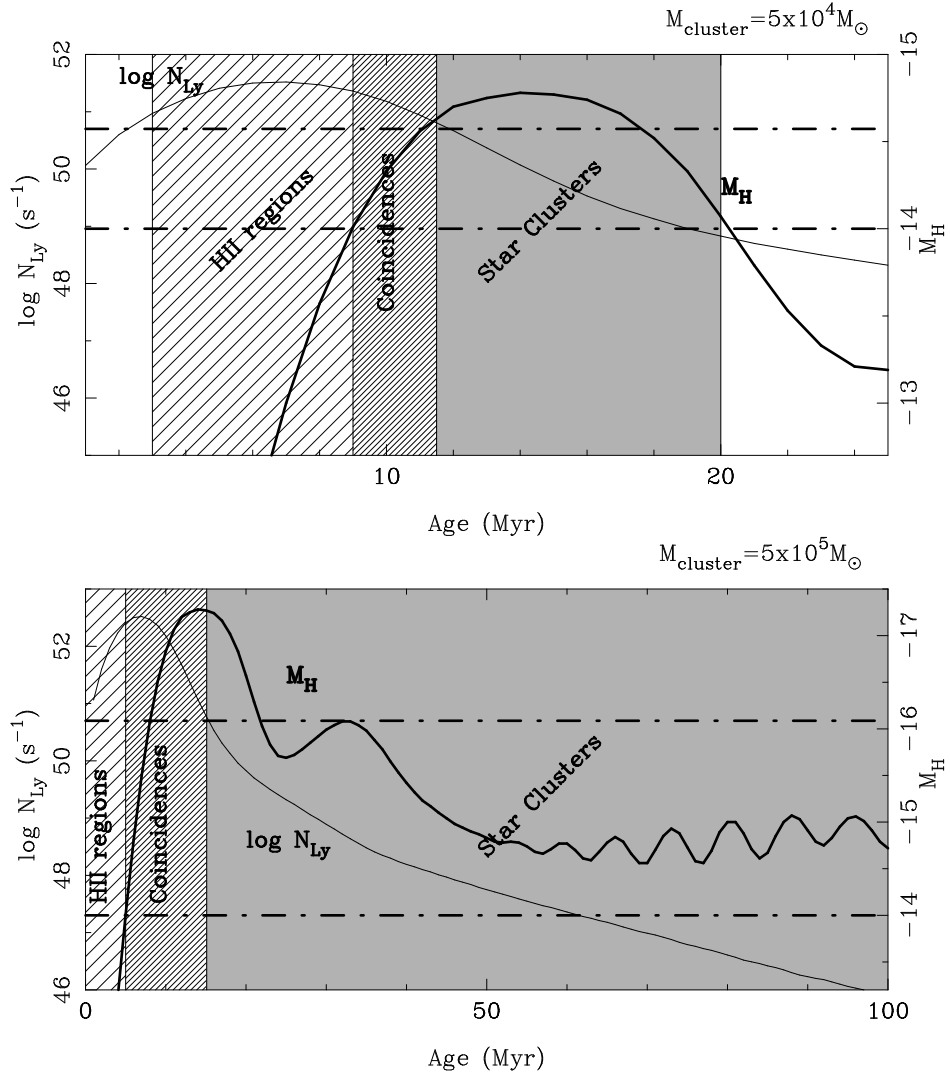}{425pt}{0}{70}{70}{-220}{30}
\vspace{-2.cm}
\caption{Same as Figure~5, but for outputs of Rieke et al. 
(1993) models, using a truncated Salpeter IMF (between 
0.1 and $80\,{\rm M}_\odot$), a Gaussian burst of 
5\,Myr duration and a cluster with $5\times 10^5\,{\rm M}_\odot$ mass
(lower panel) and $5\times 10^4\,{\rm M}_\odot$ (upper panel).}
\end{figure*}

In Figure~5 we show outputs of Starburst99 for  
two different cluster masses: $10^5\,{\rm M}_\odot$ 
and $10^6\,{\rm M}_\odot$, a Salpeter IMF (with 
lower and upper mass cutoffs of $1\,{\rm M}_\odot$ and $100\,{\rm M}_\odot$ 
respectively), instantaneous 
star formation and solar metallicity.  The solid lines represent the 
time evolution of the number of ionizing photons (thin solid line) 
and the $H$-band absolute magnitude (thick solid line). For simplicity we only 
show the detection thresholds for NGC~3256 
as dashed-dotted lines at $M_H=-14\,$mag (scale on the 
right hand side of the diagram) and $\log N_{\rm Ly}
=50.7\,{\rm s}^{-1}$ (scale on the left hand side of the diagram). 
For a $10^6\,{\rm M}_\odot$  cluster (lower panel of
Figure~5), and for ages up to $\simeq 
7\,$Myr we will observe a coincidence between an H\,{\sc ii}
region and a near-infrared star cluster (i.e., when 
both $N_{\rm Ly}$ and $M_H$ are above their 
respectively detection thresholds, shown as the closely hatched 
area), assuming no extinction (see Section~5.4.1).
After 7\,Myr and  for at least 100\,Myr, we will  
be able to detect star clusters (grey area), but no coincidences,
since $N_{\rm Ly}$ has fallen below its threshold.

The case of a Starburst99
$10^5\,{\rm M}_\odot$  cluster is displayed in the 
upper panel of Figure~5. The detection limits in NGC~3256
will allow us to identify H\,{\sc ii} regions for 
up to 5\,Myr (shown as a lightly 
hatched area) and near-infrared star clusters
for ages $7-12.5\,$Myr (shown as the grey color area). After approximately
$12-13$\,Myr,  the star clusters will be too faint to be detected 
with the current detection threshold. 
Note that for clusters of  $10^5\,{\rm M}_\odot$ mass there will be no
age for which we could detect both an H\,{\sc ii} region and 
a near-infrared star cluster (that is, there will be no coincidences). 
Within the present detection limits, for  
NGC~3256 and Arp~299 star  clusters with masses of less than  
$10^4\,{\rm M}_\odot$ will not be detected as either 
star clusters or H\,{\sc ii} regions, as at no time during 
their evolution will they be brighter than $M_H = -14\,$mag
or $\log N_{\rm Ly} = 50.9\,{\rm s}^{-1}$, respectively.

For a more extended period of star formation we have 
used outputs of Rieke et al. (1993) models with a 
Gaussian star formation rate of 5\,Myr FWHM. 
The results are presented in 
Figure~6 for two clusters with masses 
$5\times10^4\,{\rm M}_\odot$ and $5\times10^5\,{\rm M}_\odot$ 
for a truncated Salpeter IMF\footnote{$\phi(m){\rm d}m 
\propto m^{-2.35}{\rm d}m$ 
between 1 and $80\,{\rm M}_\odot$, and $\phi(m){\rm d}m \propto 
m^{-1}{\rm d}m$ between 0.1 and $1\,{\rm M}_\odot$.}. 
Figure~6 is coded in the same way as Figure~5. It is clear that 
the different star formation rates will produce different fractions of 
H\,{\sc ii} regions, coincidences and star cluster detections. For
instance, in the case of a $5\times10^5\,{\rm M}_\odot$ cluster formed 
in a Gaussian burst, we will detect H\,{\sc ii}
regions alone between 3 and 9\,Myr, before the $H$-band luminosity 
of the cluster becomes detectable.
In the case of the $5\times10^4\,{\rm M}_\odot$
cluster, because
the star formation period is more extended than in the 
instantaneous burst, it is still possible to 
have coincidences for a short period of time, between 9 and 12\,Myr. 

We now need to assume a mass distribution for the star clusters.
Elmegreen \& Efremov (1997) proposed  that all types of clusters 
(young star clusters, globular clusters, open 
clusters and associations) form with nearly 
constant efficiency in gas clouds. This 
universal mechanism applies then to all regions regardless of
epoch or geometry. Thus the  mass distribution of the star clusters 
in NGC~3256 and Arp~299 can be represented as a power law:

\begin{equation}
n(M)\,{\rm d}M \propto M^{-a}\,{\rm d}M
\end{equation}

\noindent where $a$ is the slope of the 
distribution. From star cluster and H\,{\sc ii}
region LFs, it is found that the slope 
has values of  $a = 1.7-2$ (see Elmegreen \& Efremov 1997 and 
references therein). In particular, Zepf et al. (1999) have found a 
slope of $a=1.8$ for the optical LF of 
the young star clusters in NGC~3256. 

The next step is 
to assume an age distribution for the star clusters. 
From optical spectroscopy of the H\,{\sc ii} regions of 
Arp~299, we measured a range of equivalent widths 
of H$\alpha$ (a similar situation is found in NGC~3256, see
L\'{\i}pari et al. 2000) which implies that the star clusters/H\,{\sc ii}
regions are not coeval. We will assume that clusters are formed 
at a constant rate, so there is a range of ages. Using Equation~(2) for the mass
distribution of the star clusters, and 
the evolutionary synthesis models presented above, we can 
obtain an estimate the 
fractions of young, intermediate and old clusters. For the two 
galaxies in consideration we will only need to take into account  
star clusters with masses between $\simeq 5\times 10^4\,{\rm M}_\odot$ and 
$10^6\,{\rm M}_\odot$. Less massive clusters will not be observed 
with the present detection threshold, 
whereas the upper limit of the $H$-band luminosity distribution 
($M_H\simeq -17.2\,$mag) indicates that there are no
star clusters more massive than $10^6\,{\rm M}_\odot$. 
The latter seems to be true for all the interacting/merging LIRGs in 
our sample.

\begin{deluxetable}{lcccccc}
\small
\tablewidth{16cm}
\tablecaption{Observed Numbers and Fractions of H\,{\sc ii} regions and 
near-infrared star 
clusters.}
\tablehead{\colhead{Galaxy} & \colhead{No. H\,{\sc ii}}  & 
\colhead{No. } & \colhead{No. star} & \colhead{Young H\,{\sc ii}}
& \colhead{Intermediate} & \colhead{Old}\\
&                            \colhead{regions} &  
\colhead{Coincidences}&    \colhead{clusters} &
\colhead{regions} & \colhead{clusters/H\,{\sc ii} regions} & 
\colhead{clusters}\\
(1) & (2) & (3) & (4) & (5) & (6) & (7)} 
\startdata
NGC~3256 &129 & $27-18$  & 221 & $32-33$\% & $8-5$\%  & $60-62$\% \\
IC~694   & 58 & $10-7$  & 131 & $27-28$\% & $6-4$\% & $67-68$\%\\
NGC~3690 & 75 & $25-19$ & 210 & $19-21$\% & $10-7$\% & $71-72$\%\\
NGC~1614 & ring(20) + 30 & $16-11$ & 99 & $26-28$\% & $12-8$\% & $62-64$\%\\
NGC~6808 & 81 &  $18-11$ & 122 & $34-36$\% & $10-6$\% & $56-58$\% \\
NGC~5653 & 71 & $24-17$  & 144 & $24-27$\% & $14-9$\% & $62-64$\%\\
VV~114   & 35 & $20-15$  & 52  & $22-28$\% & $30-21$\% & $48-51$\%\\
\hline
\multicolumn{7}{c}{Complete distributions}\\
\hline
NGC~3256 & 128 & $27-18$ & 170 & $37-39$\% 
& $10-7$\% & $53-54$\%\\
IC~694 &57 &$9-6$ & 99  & $33-34$\% 
& $6-4$\% & $61-62$\%\\
NGC~3690 &73 &$22-16$ & 158  & $25-27$\% 
& $10-7$\% & $65-66$\%\\
\enddata
\tablecomments{The range in numbers in columns (3), (5), (6) and 
(7) correspond to the criteria used for counting a coincidence 
between an H\,{\sc ii} region and 
a cluster: a 
separation of $2-1.5$ pixels }
\end{deluxetable}

\begin{deluxetable}{lcccc}
\small
\tablewidth{16cm}
\tablecaption{Predicted fractions of H\,{\sc ii}
regions, coincidences and old clusters for NGC~3256 and Arp~299
(no destruction of clusters).}
\tablehead{\colhead{Galaxy} & \colhead{Age}  & 
\colhead{Young H\,{\sc ii}}
& \colhead{Intermediate} & \colhead{Old}\\
 & Distribution  & \colhead{regions/clusters}   &  
 \colhead{clusters/H\,{\sc ii} regions} & \colhead{clusters}} 
\startdata
\multicolumn{5}{c}{Instantaneous burst}\\
\hline
NGC~3256 & $0-10\,$Myr & 50\% & 14\% & 36\%\\
         & $0-20\,$Myr & 36\% & 10\% & 54\%\\
         & $0-40\,$Myr & 28\% & 8\% & 64\%\\
\smallskip
         & $0-100\,$Myr& 17\% & 5\% & 78\%\\
Arp~299  & $0-10\,$Myr & 50\% & 14\% & 36\%\\
         & $0-20\,$Myr & 31\% & 9\% & 60\%\\
         & $0-40\,$Myr & 25\% & 7\% & 68\%\\
         & $0-100\,$Myr & 16\% & 4\% & 79\%\\
\hline
\multicolumn{5}{c}{$5\,$Myr Gaussian burst}\\
\hline
NGC~3256 & $0-10\,$Myr& 80\% & 20\% & 0\%\\
and Arp~299 & $0-20\,$Myr & 34\% & 23\% & 43\%\\  
         & $0-40\,$Myr & $29-27$\% & $19-18$\% & $52-55$\%\\
         & $0-100\,$Myr& $20-19$\% & $13-12$\% & $67-69$\%\\
\enddata
\end{deluxetable}

\subsection{Comparison with observations}

The first result is that the relative 
numbers of young H\,{\sc ii} regions/clusters,
intermediate and old clusters 
depend only slightly on the slope of the mass distribution (for 
the range $a = 1.7-2$). The main factors determining 
these fractions are  the age distribution of the 
star clusters (that is the age of the oldest 
clusters), and the adopted form for the IMF and 
star formation rate. See Sections 5.4 and 5.5 for a discussion on 
the effects of extinction and cluster destruction.

In Table~5 we summarize the results for a  cluster mass 
distribution with  a slope of $a =1.7$. The first part of the table are
the results for an instantaneous burst, whereas the second part are those 
for a more extended period of star formation. 
The model predictions can be compared with the fractions for the 
complete distributions of  NGC~3256 and 
Arp~299 given in the second part of Table~4. 

Inspection of  Figures~5 and 6, and 
taking into account that less massive clusters are more 
numerous than more massive clusters, shows that 
the more extended star formation activity of Rieke et al. (1993) models
will always produce more coincidences than observed, 
regardless of the maximum age of
the star clusters. This probably indicates that the instantaneous star formation 
is a better choice to account for the observed properties of 
star clusters. This seems to be the case for local star clusters where 
the age spread is at most just a few million years (e.g., Luhman et al. 1998;
Palla \& Stahler 2000).
This is not true in general for the nuclei 
of interacting galaxies, and regions of high star formation 
activity (see AAH00 and AAH01) where 
we have found that more extended periods of star formation
are required to explain the observed properties. 

Figures~5 and 6 also illustrate 
the dependence of the  cluster luminosity with the model assumptions, 
which in turn translates into different mass estimates from the
observed absolute $H$-band magnitudes.
Using the instantaneous star formation and a Salpeter IMF 
between 1 and $100\,{\rm M}_{\odot}$ we
can constrain the masses of the identified star clusters in 
NGC~3256 and Arp~299 between $\simeq 5 \times 10^4\,{\rm M}_\odot$ and 
$\simeq 10^6\,{\rm M}_\odot$. The high end of our 
photometric masses is in good agreement with 
the dynamical masses measured by Mengel et al. (2002) for compact young 
star clusters in the Antennae system.

Another interesting result from our simulations is that the observed relative 
fractions of young (H\,{\sc ii} regions) and 
old (near-infrared star clusters)  populations provide some information 
on the age spread of the last epoch of star formation. 
From the statistical point of view our simulations  
show  that most of the 
{\bf detected} stellar population in Arp~299 and NGC~3256 shows 
an age spread of between 20\,Myr and 40\,Myr, as otherwise we should 
have measured a higher fraction of old clusters (e.g., for an 
age distribution of up to 100\,Myr this faction is $\simeq 80$\%). 
Zepf et al. (1999) also deduced young ages for the NGC~3256 star clusters
detected in the optical based on the lack of a strong color-luminosity 
relation. The youth of the detected clusters does not 
exclude the presence of older clusters possibly created 
in this or 
previous episodes of star formation. These clusters 
however cannot be identified
with the present detection threshold.

\subsection{Limitations}
In this section we discuss some limitations of the 
comparison between the observed fractions of H\,{\sc ii}
regions, old clusters and the model predictions.
\subsubsection{The effects of extinction}
So far we have ignored the effects of extinction in our comparison of
observations and models. 
From detailed studies of obscured Galactic giant H\,{\sc ii} 
regions there is evidence that the extinction to the stars is 
highly variable with 
values between $A_K=0$ and $A_K=4.5\,$mag (Blum, Damineli, \& 
Conti 1999; Blum, Conti, \& Damineli 2000)
for the very young H\,{\sc ii} regions. 
If high extinction is present in the H\,{\sc ii} regions of
LIRGs, then the predicted 
fractions of H\,{\sc ii} regions, coincidences 
and older star clusters will be affected. 
Although we have no way derive the extinctions from the 
current Pa$\alpha$ observations,  
we can  obtain some estimates from optical spectroscopy. 
AAH00 and L\'{\i}pari et al. (2000) have measured
extinctions of up to $A_V \simeq 4.2\,$mag using the Balmer decrement
for a few H\,{\sc ii} regions in Arp~299 and NGC~3256 respectively.  
This is equivalent to extinctions at the observed 
wavelength of Pa$\alpha$ of up to $A_{{\rm Pa}\alpha} \simeq 0.5\,$mag. 
The ages of 
these H\,{\sc ii} regions from the observed equivalent widths of 
H$\alpha$ are of between 3 and 6\,Myr using Leitherer et al. 
(1999) models.

Although the measured extinctions are not high enough to compromise
the detection of 
H\,{\sc ii} regions at near-infrared wavelengths, we cannot rule out 
the possibility that very young H\,{\sc ii} regions ($<3\,$Myr) 
suffer from elevated extinctions, even at near-infrared 
wavelengths. If this were the case, we would be
missing the youngest  H\,{\sc ii} regions, and hence the observed
fraction of H\,{\sc ii} regions and coincidences will be lower limits. This
would in turn translate in even younger age 
distribution of the detected population 
of star clusters. 

\subsubsection{The effects of destruction of clusters}
The observed luminosity and 
mass functions of old globular clusters around galaxies and young 
star clusters observed in interacting 
galaxies appear to be distinctively different, with the former having a 
log-normal shape, and the latter a power law form with no 
evidence for a turnover (see for instance
Elmegreen \& Efremov 1997; Whitmore et al. 1999; Zepf et al. 1999).
This poses an interesting problem if the clusters observed in 
galaxies  
are younger versions of today's globular clusters. One of the proposed
solutions is that the mass distribution of old globular clusters was initially 
a power law that was later modified by selective destruction of 
low mass star clusters to become a log-normal like mass function
(see Fall \& Zhang 2001 for a detailed 
discussion, and also Whitmore 2000).

Although including a detailed treatment of the problem of cluster destruction 
is beyond the scope of this paper, we can attempt to see its effects on our 
calculations.  If clusters in Arp~299 and NGC~3256 
have been forming at a constant 
rate for approximately $100\,$Myr, then $\simeq 50\%$ of all clusters 
with masses above $5\times 10^4\,{\rm M}_\odot$ 
should  be destroyed during that period to account for the observed 
fraction of star clusters. Note that Zepf et al. (1999) proposed 
selective destruction of low mass clusters as
one possibility to account for the observed optical colors and luminosities
of clusters in NGC~3256 (the other possibility was a very young 
age for the clusters). 

\subsubsection{Pa$\alpha$ shells}
Finally we consider the effects of the formation of 
Pa$\alpha$ shells on the 
measured number of coincidences. In the Antennae galaxy there is
evidence that many of the slightly older, more massive clusters 
(i.e., 5-10\,Myr) have blown large H$\alpha$ shells around themselves, 
hence there is
no longer a good correspondence between the distribution of H$\alpha$
and the $I$-band center of the cluster (Whitmore et al. 1999).
Such Pa$\alpha$ shells
are diffuse and will 
not be identified as H\,{\sc ii} regions, causing the number  
of coincidences to be underestimated in the $5-7\,$Myr age range.
For ages $>7\,$Myr the H\,{\sc ii} region emission will not be
detected as its luminosity will be below our detection threshold
(Section~5.2). The result of missing some 
coincidences because of the presence of Pa$\alpha$ shells would be  
an age distribution of the clusters slightly younger than inferred
in Section~5.3. Note that this effect is 
only relevant to the most massive clusters in Arp~299 and 
NGC~3256 where we expect to see coincidences with the 
present detection threshold. 

\section{Discussion and Conclusions}
In this paper we have presented {\it HST}/NICMOS 
broad-band and narrow-band Pa$\alpha$ imaging of a sample of 8 LIRGs. 
The sample galaxies exhibit a range of infrared luminosities
($\log L_{\rm IR} = 10.94-11.82\,{\rm L}_\odot$), as well as a variety 
of dynamical stages: isolated galaxies interacting galaxies and mergers.
The Pa$\alpha$ images have allowed us to identify the location of 
H\,{\sc ii} regions, whereas the $H$-band continuum images have revealed
the star clusters. In all galaxies in our sample except NGC~6240 and 
Zw~049.057 we have detected a large number of star clusters and 
H\,{\sc ii} regions.  The absolute $H$-band magnitudes of the 
identified SSCs range up to $M_H=-17.2\,$mag. A large  
fraction of the  H\,{\sc ii} region population (excluding the nuclear
emission) shows luminosities in excess of that of 30 Doradus, the prototypical
giant H\,{\sc ii} region.

The main characteristic of the Pa$\alpha$ emission in the isolated LIRGs
in our sample is the lack of strong nuclear emission. Most of the 
 H\,{\sc ii} regions are distributed along the spiral arms of the galaxies.
The interacting/merging LIRGs on the other hand, show bright nuclear
Pa$\alpha$ emission together with widely spread star formation along the 
the spiral arms and at the interface of interacting galaxies. The fraction
of nuclear Pa$\alpha$ emission to the total emission varies from system 
to system, as it depends on factors such as the age of the interaction 
process, the initial gas content of the galaxies and the relative 
masses of the galaxies. 

We have analyzed the properties -- luminosities, sizes and luminosity 
functions -- of H\,{\sc ii}
regions in LIRGs at an unprecedented spatial resolution (between 
15 and 78 pc). Giant H\,{\sc ii} regions 
are ubiquitous in LIRGs and are located not only near 
the nuclei of interacting galaxies, but also 
at the interface of interacting galaxies and along 
the spiral arms.  This population of highly luminous H\,{\sc ii}
regions is not observed in normal galaxies. 
We have fitted power laws to the H\,{\sc ii} region LFs of 
Arp~299 and NGC~3256 and found that the indices 
are within the values previously measured
in the disks of normal galaxies. 

We have compared the properties of the 
H\,{\sc ii} regions in LIRGs with a small sample of normal 
galaxies observed with the same spatial resolution and found 
that giant H\,{\sc ii} regions are more common in LIRGs than in 
normal galaxies. The measured sizes
of giant H\,{\sc ii} regions in LIRGs compared to those in normal galaxies 
rule out the possibility that the giant H\,{\sc ii} regions are just
aggregations of ``normal'' H\,{\sc ii} regions. Another possibility 
would be that all H\,{\sc ii} regions in LIRGs were extremely young. However,
previously published spectroscopy of a few 
giant H\,{\sc ii} regions in Arp~299 and NGC~3256 has
shown a range of equivalent widths of H$\alpha$, in other words,  a range 
of ages. A more plausible explanation for the presence of
extremely luminous H\,{\sc ii}
regions in LIRGs is that the regions of high gas pressure and 
density present in  LIRGs, ULIRGs and interacting galaxies 
provide the necessary conditions for the formation of a 
large number of massive star (ionizing) clusters. Such extreme 
conditions are not likely to occur in 
normal isolated galaxies.

Despite the large numbers of near-infrared 
SSCs and H\,{\sc ii} regions identified in LIRGs, there is 
only a small  
fraction of coincidences -- between 4\% and 30\% of the total number of
detected H\,{\sc ii} regions and star clusters. This is suggestive 
of an evolutionary sequence similar to that seen in obscured Galactic
giant H\,{\sc ii} regions. For the first 
few million years of the cluster evolution, there will be 
significant amounts of dust which may preclude the detection of the 
ionizing cluster. As the cluster ages, stellar winds and supernovae
will dissipate part of the gas and dust, and the star cluster will 
become visible, whereas at the same time the H\,{\sc ii} region 
emission will be less luminous.

We have used evolutionary synthesis models to reproduce the 
observed  relative fractions of young 
and intermediate H\,{\sc ii} regions/clusters and old clusters 
in Arp~299 and NGC~3256. Based on the
observed fractions we have concluded that most likely the star 
formation occurs in instantaneous bursts rather than more extended 
periods of star formation. Using instantaneous star 
formation and a Salpeter IMF we have 
derived photometric masses of the detected star clusters
of between $5\times 10^4\,{\rm M}_{\odot}$ 
and $10^6\,{\rm M}_{\odot}$.

The fact that the peak of the $H$-band 
luminosity occurs after approximately 9\,Myr whereas at the 
same time the number of ionizing photons has dropped by 
about two orders of magnitudes from the maximum 
explains the limited number of coincidences. Within the 
present detection limits in Arp~299 and NGC~3256, 
we can only detect both H\,{\sc ii} region 
emission and a star cluster for the most massive clusters 
($\simeq 10^6\,{\rm M}_{\odot}$) during the  
first 7\,Myr. The near-infrared  clusters with no detected 
H\,{\sc ii} region emission will be older than approximately 7\,Myr. 
The H\,{\sc ii} regions with no detected  cluster 
counterpart are most likely younger than 
5\,Myr, and have intermediate mass ($\simeq 5\times10^4-10^5\,{\rm M}_\odot$)
ionizing clusters. If, as observed in 
obscured Galactic H\,{\sc ii} regions, there are significant 
amounts of extinction during 
the first million years of the evolution of clusters and 
H\,{\sc ii} regions,  then we may be
missing the youngest star forming regions, and hence the observed
fractions of H\,{\sc ii} regions and coincidences will be lower limits.

An estimate of the age distribution of the {\bf observed} 
clusters can be inferred from 
the relative numbers of H\,{\sc ii} regions, and near-infrared 
star clusters: the higher the fraction of near-infrared clusters, 
the older the ages of the detected star clusters will be. 
We find that the ages of the detected 
star clusters in Arp~299 and NGC~3256 range up to $20-40$\,Myr.
Older clusters possibly created in this or 
previous episodes of star formation are likely to 
exist in these systems but cannot be identified
with the present detection threshold. Another possibility to 
explain the apparent youth of the clusters in Arp~299 and NGC~3256 would 
be destruction of clusters. In that case, if the clusters have been created 
at a constant rate for the last 
$\simeq 100\,$Myr, then roughly 50\% of the clusters 
are destroyed 
during that time to account for the observed 
fraction of clusters in these two systems. 
The data presented in this paper does not
allow us to distinguish between these two possibilities.

From the present observations 
and modelling it is clear that a large population of the 
youngest clusters (that is the ionizing clusters of the  
H\,{\sc ii} regions with ages of $<5-6\,$Myr) will not be 
detected from near-infrared continuum imaging alone, as only some $8-16\%$ of 
these H\,{\sc ii} regions in our sample of LIRGs (both isolated and 
interacting/merging systems) appear to have  
near-infrared cluster counterparts. This suggests 
that studies of the young star clusters in galaxies performed 
using only near-infrared continuum imaging may be  biased against the 
youngest star forming regions, which may account for 
up to  $40\%$ of the  detected population of star clusters 
and H\,{\sc ii} regions in LIRGs.

\section*{Acknowledgments}
It is a pleasure to thank Valentin Ivanov and John Black for 
enlightening discussions on the paper. We are also grateful to an anonymous
referee for helpful comments, which resulted in an improved paper.

This work has been partially supported by the National
Aeronautics and Space Administration grant NAG 5-3042 
through the University of Arizona  and 
Contract 960785 through the Jet Propulsion Laboratory.

This research has made use of the NASA/IPAC Extragalactic Database (NED) 
which is operated by the Jet Propulsion Laboratory, California Institute 
of Technology, under contract with the National
Aeronautics and Space Administration.

\end{document}